\documentclass[amsmath,amssymb,aps,pra,preprint,showpacs,superscriptaddress]{revtex4-1}
\usepackage{graphicx}
\usepackage{dcolumn}
\usepackage{bm}
 \usepackage{color}
\begin{document}
\title{Generation and nonclassicality of entangled states via the interaction of two three-level atoms with a quantized cavity field assisted by a driving external classical field}
\author{H. R. Baghshahi}
\affiliation{Atomic and Molecular Group, Faculty of Physics, Yazd University, Yazd, Iran}
\affiliation{The Laboratory of Quantum Information Processing, Yazd University, Yazd, Iran}
\affiliation{Department of Physics, Faculty of Science, Vali-e-Asr University of Rafsanjan, Rafsanjan, Iran}
\author{M. K. Tavassoly}
\email[]{mktavassoly@yazd.ac.ir}
\affiliation{Atomic and Molecular Group, Faculty of Physics, Yazd University, Yazd, Iran}
\affiliation{The Laboratory of Quantum Information Processing, Yazd University, Yazd, Iran}
\author{S. J. Akhtarshenas}
\affiliation{Department of Physics, Ferdowsi University of Mashhad, Mashhad, Iran}
\date{\today}

\begin{abstract}
The interaction of two identical three-level atoms of the types $V$, $\Xi$ and $\Lambda$ with a quantized cavity field as well as a driving external classical field is studied.  Under two certain unitary transformations, the system is converted to a typical form of the Jaynes-Cummings model for two three-level atoms. The exact analytical solutions of the wave function for different considered atom-field systems are exactly obtained with the help of the Laplace transform technique, when the atoms are initially prepared in the topmost excited state and the quantized field is in a coherent state. In order to examine the nonclassicality features of the deduced states, the dynamics of the entanglement between subsystems is discussed via two well-known measures, namely, von Neumann entropy of the reduced state and negativity. In addition, we pay attention to the temporal behaviour of quantum statistics of the photons of the field and squeezing phenomenon. Meanwhile, the influence of the external classical field on the latter physical quantities is analyzed in detail. The results show that the mentioned quantities can be sensitively controlled via the external classical field. Also, numerical computations imply the fact that the nonclassicality features in $\Xi$-type three-level atomic system is more visible than the other two configurations. In addition, it is shown that in the particular case of $\Lambda$-type atomic system, the rank of the reduced density matrix of the three-level atoms is no larger than three, so that negativity fully captures the entanglement of this system and that such entanglement is distillable.
\end{abstract}
\maketitle
  \section{Introduction}
 \label{intro}
The light-matter interaction is an essential concern in optical physics. A simple paradigm of this interaction contains a two-level atom coupled to a single-mode quantized radiation field in an optical cavity. Whenever the strength of the atom-field coupling  is far smaller than the field frequency, the rotating wave approximation (RWA) is applicable and the system is described by the well-known Jaynes-Cummings model (JCM) \cite{Jaynes.Cummings1963,Cummings1965}.
The simplicity of this model and its potential applications for more complicated and generalized atom-field systems, together with revealing some non-trivial phenomena and extraordinary characteristics, such as the collapses and revivals of Rabi oscillations in atomic population inversion  \cite{Shore.Knight1993}, placed the JCM at the heart of this area of research in quantum optics.
 Experimentally, this model can be realized when atoms are coupled to a nanomechanical oscillator \cite{LaHaye2009} and nuclear spins interacting with a magnetic field \cite{Irish2003}.
Many generalizations of the JCM have been proposed in various ways, for instance, considering different initial conditions \cite{Kuklifmmodecutenlseniski.Madajczyk1988}, entering the effects of dissipation and damping in the model \cite{rodriguez2005combining}, considering intensity-dependent coupling \cite{Buck.Sukumar1981,Buck.Sukumar1984,Faghihi.Tavassoly2013,Baghshahi.Tavassoly2014}, adopting a multi-level atom \cite{Faghihi2014,Faghihi2014a,BaghshahiC2014} as well as a multi-photon transition \cite{Baghshahi.etal2014} and a multi-atom \cite{Bougouffa2013,Hosny2011}.\\
From another perspective of this area of research, in recent decades, a lot of attention has been paid to the study of different configurations of three-level
atoms ($V$, $\Xi$ and $\Lambda$ type) .
 For example, there exist many theoretical works containing the interaction between a three-level atom and a single-mode cavity field \cite{Ashraf1994,Zait2003,Obada2012,Obada.etal2014}. In some of them, some nonclassicality features have been examined by considering the resonance condition between atomic transitions and quantized field frequency.
 The nonclassical properties of three-level atomic systems have been well investigated in order to understand the quantum coherence phenomena such as electromagnetically induced transparency (EIT) \cite{Boller1991}, lasing without inversion \cite{Scully1989}, and coherent trapping \cite{Scully.Zubairy1997}. In detail, the $V$-type three-level atoms are extensively used in studying nonclassicality features such as quantum beats \cite{Scully.Zubairy1997}, quantum Zeno effect \cite{Chiu1977} and quantum jumps \cite{Cook1985}. The application of $\Xi$-type atoms is well established in the coherent population trapping \cite{Tajalli2003} and also the experiments which are designed to achieve laser cooling in trapped ions \cite{Marzoli1994}. Also, the $\Lambda$-type atomic systems have been widely utilized in representing the coherent phenomena such as EIT \cite{Boller1991} and stimulated Raman adiabatic passage \cite{Bergmann1998}.\\
In particular and in direct relation to the present work, the dynamics of a physical system governed by the JCM can lead to the highly atom-field entangled state, almost as though the atom and the field can form some sort of ``molecule'' \cite{Alsing.etal1992}; the case that is obviously revealed through the vacuum Rabi splitting \cite{Sanchez-Mondragon1983}. It may be noted that, although, such experiment regarding the observation of a molecule is not usually a routine work, however,
this may be naturally probed through an external field.
Furthermore, strong driving fields have shown other attractive applications in the atom-field interaction such as detection of Fock states of the radiation field \cite{Varcoe2000},  quantum-phase gate \cite{Turchette1995}, generation of the multi-partite entangled states \cite{Rauschenbeutel1999,Solano2003,Miryp2014} and controlling nonclassical properties of system \cite{Li2000}. Henceforth, the driving JCM (DJCM), that is, the JCM which contains an external driving field, has attracted much attention in recent decades \cite{Dutra1994,Chough1996,Gerry2002,Akhtarshenas2010} .
The model for a two-level atom interacting with external quantum and classical electric fields through a  parametric frequency converter has been presented in \cite{Abdalla2010}. In this attempt, the system is detracted to an effective JCM by adequate adjustment of the field coupling in the frequency converter. In addition, it has been shown that the dynamics of physical quantities related to the atom and the atom-photon entanglement can be controlled by the classical field.
The implementation of a strongly driven one-atom laser, based on the off-resonant interaction of a $\Lambda$-type three-level atom with a single-mode cavity field and three laser fields have been schematically proposed in \cite{Lougovski2007}. The authors showed that the system can be equivalently described by a two-level atom resonantly coupled to the cavity field assisted by a strong effective coherent field. Altogether, in all above discussions, it seems that the driving external field is a suitable parameter for controlling the nonclassicality features, especially entanglement of the atom-field systems.\\
In this paper, we aim to study the problem of two identical three-level atoms in three different types ($V$, $\Xi$ and $\Lambda$ configurations) interacting with a single-mode cavity field in the presence of an external classical field through which new classes of entangled state can be generated. In working with the mentioned configurations, we are able to transform the interaction formalism to the generalized JCM by applying two appropriate unitary transformations. The explicit form of the entangled state vector of the whole system can then be exactly obtained by using the time-dependent Schr\"{o}dinger equation with the help of the Laplace transform technique. In this respect, we take the atoms to be prepared in their higher excited states and the field is supposed to be in a coherent state. Then, briefly the main goal of this paper is to investigate the effect of external classical field on the entanglement dynamics between subsystems and some of the well-known nonclassicality features. To achieve this purpose, the degree of entanglement (DEM) between subsystems through von Neumann entropy of the reduced state (to investigate the atom-field entanglement) and negativity (to obtain the atom-atom entanglement) are numerically studied. Also, the time evolution of quantum statistics and squeezing are examined in detail. We show that the dynamics of the mentioned physical quantities can be tuned by the strength of the external classical field.\\
In order to give more explanation about our motivations, it is instructive to give a few words regarding the significance and the notability of the considered systems containing the three-level atoms. In quantum information processing (QIP) the three-level systems possess outstanding advantages in comparison with two-level ones \cite{Bechmann-Pasquinucci2000a}. In this regard, the optimal eavesdropping in quantum cryptography with three-dimensional systems has been studied in \cite{Bruss2002} in which the authors found that the three-dimensional scheme offers higher security than the two-dimensional systems.
Kaszlikowski \textit{et al} investigated the general case of two entangled quantum systems defined in $d$-dimensional  Hilbert spaces, or `qudits' and have shown that violations of local realism are stronger for two maximally entangled qudits ($3\leq d \leq9$) than for two qubits \cite{Kaszlikowski2000}.
Also, based on the Greenberger-Horne-Zeilinger (GHZ) theorem, the conflict of local realism and quantum mechanics for three or more qubits has been reported much sharper than for two qubits \cite{Kafatos2010}. In this case, there exist some suggestions which indicate the reduction of this conflict due to increasing $d$ (dimension of Hilbert space) \cite{Mermin1980,Garg1982,Ardehali1991}. So, it seems that $d$-level quantum systems, or qudits,  may be better candidates to be utilized in the theoretical/experimental observations. The mainspring of this usage is to increase the available Hilbert space with the same amount of physical resources \cite{Greentree2004}. Accordingly, three-level atoms have received noticeable attention in the studies that concern with the atom-field interaction \cite{Dutta2006,Faghihi.Tavassoly2012,Faghihi.Tavassoly2013a,Faghihi.etal2013}.
 In particular and in direct relation to the considered model in this paper, quantum information processing using superconducting qubits has made outstanding advances in the past few years \cite{Makhlin2001,You2006}. One-qubit and two-qubit quantum circuits have been realized experimentally in superconducting systems. One of the most important issues in quantum information processing is how to couple two qubits, which has been widely studied theoretically and experimentally in superconducting quantum circuits. Theoretical proposals have been put forward to selectively couple any pair of qubits through a common data bus (LC circuit or a cavity field) \cite{You2001,You2002}. Liu \textit{et al.} experimentally presented a proposal to achieve a controllable interaction between flux qubit by virtue of time-dependent electromagnetic field \cite{Liu2006}. Also, a coupling (decoupling) method between a superconducting qubit and a data bus that uses a controllable time-dependent electromagnetic field (TDEF) have been studied theoretically in \cite{Liu2006a}. This reference indicates that, by choosing appropriate parameters for the TDEF, the dressed qubit (qubit plus the electromagnetic field) can be coupled to the data bus and, thus, the qubit and the data bus can exchange information with the assistance of the TDEF. The superconducting qubit circuits generalized to superconducting qudit circuits for more complex quantum computational architectures, and for richer simulations of quantum mechanical systems \cite{Neeley2009}. The considered model in this paper may be supposed as a superconductor with two qutrits (three-level atoms), LC circuit (single-mode cavity field) and TDEF (the classical field). In this case, the entanglement between two qutrits as well as between two qutrits and a single-mode field can be controlled by external classical field.\\
The reminder of the paper is structured as follows. In the next section, the model containing all existing interactions is introduced and then by applying two distinct unitary transformations, the model is reduced to the typical form of the generalized JCM. In section \ref{The entire state vector of the system}, the state vector of the whole atom-field systems is analytically obtained. Section \ref{Entanglement} deals with examining the effect of external classical field on the DEM between subsystems via von Neumann reduced entropy and negativity. In addition, in order to study the nonclassicality features of the obtained states, Mandel parameter and quadrature squeezing are respectively investigated in sections \ref{Photon statistics: the Mandel parameter} and \ref{Squeezing}. Finally, the main results of the paper are summarized in section \ref{Conclusion}.
 \section{Description of the model}\label{model}
 We consider two identical three-level atoms (labeled by $A$ and $B$) in three different configurations (namely $V,$ $\Xi$ and $\Lambda$ types as depicted in figure 1), with states $|1\rangle$, $|2\rangle$, and $|3\rangle$ and their corresponding energies $\omega_{1}$, $\omega_{2}$, and $\omega_{3}$. The atomic system is driven by an  external classical field with frequency $\omega_{c}$, and is coupled to a single-mode quantized radiation field with frequency $\omega_{f}$. Accordingly, the total Hamiltonian can be appropriately described by
\begin{eqnarray} \label{Phy 1}
\hat{H}&=& \sum_{j=A,B}\sum_{i=1}^{3}\omega_{i}\hat{\sigma}_{ii}^{(j)}+\omega_{f} \hat{a}^{\dag} \hat{a}+g_{1}\sum_{j=A,B}(\hat{\sigma}_{1}^{(j)} \hat{a}+\hat{\sigma}_{1}^{\dagger^{(j)}} \hat{a}^{\dag})\nonumber\\&+&g_{2}\sum_{j=A,B}(\hat{\sigma}_{2}^{(j)} \hat{a}+\hat{\sigma}_{2}^{\dagger^{(j)}} \hat{a}^{\dag})+\lambda_{1}\sum_{j=A,B}(\hat{\sigma}_{1}^{(j)}e^{-i\omega_{c}t}+\hat{\sigma}_{1}^{\dagger^{(j)}}e^{i\omega_{c}t})\nonumber\\&+&
\lambda_{2}\sum_{j=A,B}(\hat{\sigma}_{2}^{(j)}e^{-i\omega_{c}t}+\hat{\sigma}_{2}^{\dagger^{(j)}}e^{i\omega_{c}t}),
\end{eqnarray}
where $\hat{\sigma}_{ii}=|i\rangle \langle i|$ is the atomic projection operator,  $\hat{a}$ and $\hat{a}^{\dag}$ are respectively the bosonic
annihilation and creation operators of the field, $g_{i}$ and $\lambda_{i}$, $i=1,2$ represent the coupling constants of the interaction of the atoms with the quantized radiation and with the classical driving fields, respectively. Also, the values of $(\hat{\sigma}_{1},\hat{\sigma}_{2})$ for the three configurations are given by $(\hat{\sigma}_{13},\hat{\sigma}_{23})_{V}$, $(\hat{\sigma}_{12},\hat{\sigma}_{23})_{\Xi}$ and $(\hat{\sigma}_{12},\hat{\sigma}_{13})_{\Lambda}$.
In the rotating reference frame with frequency $\omega_{c}$ and under a unitary transformation $\hat{U}_{1}(t)=\exp[-i\omega_{c}t(\hat{a}^{\dag} \hat{a}+\sum_{j=A,B}(\hat{\sigma}_{1}^{'(j)}+\hat{\sigma}_{2}^{'(j)}))] $, the above Hamiltonian can be transformed to
\begin{eqnarray} \label{Phy 3}
\hat{H}_{1}&=&\hat{U}_{1}^{\dag}(t) \hat{H} \hat{U}_{1}(t)-i\hat{U}_{1}^{\dag}(t)\frac{d\hat{U}_{1}(t)}{dt}\nonumber\\
&=&\Delta_{1} \sum_{j=A,B}\hat{\sigma}_{1}^{'j}+\Delta_{2}\sum_{j=A,B} \hat{\sigma}_{2}^{ 'j}+\Delta \hat{a}^{\dag} \hat{a}\nonumber\\&+&g_{1}\sum_{j=A,B}(\hat{\sigma}_{1}^{(j)} \hat{a}+\hat{\sigma}_{1}^{\dagger^{(j)}} \hat{a}^{\dag})+g_{2}\sum_{j=A,B}(\hat{\sigma}_{2}^{(j)} \hat{a}+\hat{\sigma}_{2}^{\dagger^{(j)}} \hat{a}^{\dag})\nonumber\\&+&\lambda_{1}\sum_{j=A,B}(\hat{\sigma}_{1}^{(j)}+\hat{\sigma}_{1}^{\dagger^{(j)}})+
\lambda_{2}\sum_{j=A,B}(\hat{\sigma}_{2}^{(j)}+\hat{\sigma}_{2}^{\dagger^{(j)}}),
\end{eqnarray}
where the values $(\hat{\sigma}_{1}^{'},\hat{\sigma}_{2}^{'})$ of the three-level atom are in the form $(\hat{\sigma}_{11}, \hat{\sigma}_{22})$, $(\hat{\sigma}_{11}, -\hat{\sigma}_{33})$ and $(-\hat{\sigma}_{22}, -\hat{\sigma}_{33})$, respectively for $V$-, $\Xi$- and $\Lambda$-type atoms. The parameter $\Delta=\omega_{f}-\omega_{c}$ is the detuning parameter between the cavity field and the classical driving field. Also, the detuning parameters between the classical field and the atoms are given by $\Delta_{1}=(\omega_{1}-\Omega_{1})-\omega_{c}$ and $\Delta_{2}=(\Omega_{2}-\omega_{3})-\omega_{c}$, in which the values of $(\Omega_{1},\Omega_{2})$ in three configurations read as $(\omega_{3},\omega_{2})_{V}$, $(\omega_{2},\omega_{2})_{\Xi}$ and $(\omega_{2},\omega_{1})_{\Lambda}$. It is seen that by applying the mentioned transformation, the time-dependent exponential terms in the Hamiltonian (\ref{Phy 1}) have been clearly eliminated.
Now, for simplicity, let us follow the problem in the resonance conditions. Also, without loss of generality, we suppose $g_{1}= g_{2}=g$ and $\lambda_{1}= \lambda_{2}=\lambda$. Under these assumptions, we can recast the Hamiltonian (\ref{Phy 3}) as follows
 \begin{eqnarray} \label{Phy 5}
\hat{H}_{1}&=&g \sum_{j=A,B}(\hat{\sigma}_{1}^{(j)} \hat{a}+\hat{\sigma}_{2}^{(j)} \hat{a}+\hat{\sigma}_{1}^{\dagger^{(j)}} \hat{a}^{\dag}+\hat{\sigma}_{2}^{\dagger^{(j)}} \hat{a}^{\dag})\nonumber\\&+&\lambda\sum_{j=A,B}(\hat{\sigma}_{1}^{(j)}+
\hat{\sigma}_{2}^{(j)}+\hat{\sigma}_{1}^{\dagger^{(j)}}+\hat{\sigma}_{2}^{\dagger^{(j)}}).
\end{eqnarray}
 If $\lambda=0$, then Eq. (\ref{Phy 5}) describes the generalized JCM for the interaction between two three-level atoms and a single-mode quantized cavity field, while the case $\lambda\neq0$ corresponds to generalized DJCM for two three-level atoms. Now, in order to analyze the dynamics of the considered system with the Hamiltonian (\ref{Phy 5}), we utilize the probability amplitude method. Altogether, it is not still an easy work to solve the above system analytically. This is due to the existence of the external classical field in addition to the terms which are connected to the interaction between two three-level atoms and single-mode quantized field. Therefor, before using this approach, we try to reduce  the Hamiltonian (\ref{Phy 5}) to the typical form of the generalized JCM for two three-level atoms. This goal will be achieved by introducing the following displacement operator
 \begin{equation} \label{Phy 6}
 \hat{D}(\gamma)=\exp(\gamma \hat{a}^{\dag} - \gamma ^{\ast} \hat{a}),\hspace{0.75cm}  \gamma=\frac{\lambda}{g},
  \end{equation}
which satisfies the identity
  \begin{equation} \label{Phy 7}
  \hat{D}(\gamma) \hat{a}  \hat{D}^{\dag}(\gamma)=\hat{a}- \gamma.
  \end{equation}
 In this case, by applying the unitary operator given in (\ref{Phy 6}), the Hamiltonian (\ref{Phy 5}) is converted to the Hamiltonian of the form
  \begin{eqnarray} \label{Phy 8}
  \hat{H}_{2}&=&\hat{D}(\gamma) \hat{H}_{1} \hat{D}^{\dag}(\gamma)=g \sum_{j=A,B}(\hat{\sigma}_{1}^{(j)} \hat{a}+\hat{\sigma}_{2}^{(j)} \hat{a}+\hat{\sigma}_{1}^{\dagger^{(j)}} \hat{a}^{\dag}+\hat{\sigma}_{2}^{\dagger^{(j)}} \hat{a}^{\dag}).
  \end{eqnarray}
 Briefly, up to now we used the two transformations (the local transformation $\hat{U}_{1}(t)$ and the displacement operator $\hat{D}(\gamma)$) for reducing the Hamiltonian   (\ref{Phy 1}) to the typical generalization of the JCM for two three-level atoms. It is worth noticing that, the physical properties as well as the initial conditions of any system are preserved under the local transformation. But the second transformation (displacement operator) changes the initial condition and the physical features of the system. Henceforth, by using the Hamiltonian (\ref{Phy 8}) and defining the state vector $|\psi_{2}(t)\rangle=\hat{D}(\gamma)|\psi(t)\rangle$, one may consider the time-dependent Schr\"{o}dinger equation corresponding to $|\psi_{2}(t)\rangle$ as $  i \partial |\psi_{2}(t)\rangle / \partial t=\hat{H}_{2}|\psi_{2}(t)\rangle$, from which one easily obtain $|\psi(t)\rangle=\hat{D}^{\dagger}(\gamma)| \psi_{2}(t)\rangle$. In the next section, we are going to evaluate the state vector $| \psi_{2}(t)\rangle$ for different configurations of the three-level atoms.

 \section{The entire state vector of the system}\label{The entire state vector of the system}
 The main goal of this section is to obtain the state vector of the considered systems. Before achieving this purpose, it is necessary to determine the initial conditions of the atoms as well as the field. We consider the field to be initially in the coherent state and suppose that the atoms enter to the cavity in the upper exited state, i.e.,
  \begin{equation} \label{Phy 11}
| \psi(t=0)\rangle=| 1,1\rangle |\alpha\rangle,\hspace{.25cm} |\alpha\rangle=\exp \left(-\frac{|\alpha|^{2}}{2}\right)\sum_{n=0}^{\infty}\frac{\alpha^{n}}{\sqrt{n!}}| n\rangle.
  \end{equation}
 Hence, the state vector  $|\psi_{2}(t=0)\rangle$ may be obtained by the following relation
  \begin{eqnarray} \label{Phy 12}
| \psi_{2}(t=0)\rangle&=&\hat{D}(\gamma)| 1,1\rangle |\alpha\rangle=| 1,1\rangle \hat{D}(\gamma) \hat{D}(\alpha)| 0\rangle=| 1,1\rangle| \beta\rangle,
  \end{eqnarray}
 where we have used the identity
  \begin{equation} \label{Phy 121}
\hat{D}(\gamma) \hat{D}(\alpha)=\exp(i\Im(\gamma \alpha^{\ast}))\hat{D}(\beta),
  \end{equation}
  so that $\beta=\alpha+\gamma$.
  The factor $\exp(i\Im(\gamma \alpha^{\ast}))$ is dropped because the parameters $\gamma$ and $\alpha$ have been supposed to be real. By considering this initial condition for $|\psi_{2}(t)\rangle$, the wave function for the considered systems are obtained separately, in the next subsections.

  \subsection{Two $V$-type three-level atoms}\label{Two $V$-type}
  The wave function $|\psi_{2}(t)\rangle$ at any time $t$ for two $V$-type three-level atoms may be written as,
   \begin{eqnarray} \label{Phy 13}
| \psi_{2}(t)\rangle_{V}&=& \sum_{n=0}^{\infty} \bigg(C_{1}(n,t)|1,1,n \rangle +C_{2}(n,t)\big(|1,2,n \rangle+|2,1,n \rangle \big)\nonumber\\&+&C_{3}(n+1,t)\big(|1,3,n+1 \rangle+|3,1,n+1 \rangle \big)\nonumber\\&+&C_{4}(n+1,t)\big(|2,3,n+1 \rangle+|3,2,n+1 \rangle\big)\nonumber\\&+&C_{5}(n,t)|2,2,n \rangle+C_{6}(n+2,t)|3,3,n+2\rangle \bigg),
  \end{eqnarray}
  where the coefficients $C_{i}, i=1,2...6$, are the unknown probability amplitudes that should be determined. By inserting the assumed wave function (\ref{Phy 13}) into the time-dependent Schr\"{o}dinger equation together with considering the Hamiltonian (\ref{Phy 8}), the following coupled differential equations for the probability amplitudes may be found:
  \begin{eqnarray} \label{Phy 14}
\frac{dC_{1}(n,t)}{dt}&=&-iV_{1}C_{3}(n+1,t),\nonumber\\
\frac{dC_{2}(n,t)}{dt}&=&-iV_{1} \Bigg(C_{3}(n+1,t)+C_{4}(n+1,t) \Bigg),\nonumber\\
\frac{dC_{3}(n+1,t)}{dt}&=&-iV_{1}\Bigg (C_{1}(n,t)+C_{2}(n,t) \Bigg)-iV_{2}C_{6}(n+2,t),\nonumber\\
\frac{dC_{4}(n+1,t)}{dt}&=&-iV_{1}\Bigg (C_{2}(n,t)+C_{5}(n,t) \Bigg)-iV_{2}C_{6}(n+2,t),\nonumber\\
\frac{dC_{5}(n,t)}{dt}&=&-2iV_{1}C_{4}(n+1,t),\nonumber\\
\frac{dC_{6}(n+2,t)}{dt}&=&-2iV_{2}\Bigg (C_{3}(n+1,t)+C_{4}(n+1,t)\Bigg),
  \end{eqnarray}
where $V_{1}= V(n+1)$ and $V_{2}= V(n+2)$ and $V(n)= g\sqrt{n}$. After some lengthy calculations, we obtain the probability amplitudes via the Laplace transform techniques as below:
\begin{subequations}
 \begin{eqnarray} \label{Phy 15}
C_{1}(n,t)&=&\frac{1}{4}C_{1}(n,0)\Bigg(2\cos(\sqrt{2} V_{1} t)+\frac{V_{1}^{2}+2V_{2}^{2}+V_{1}^{2}\cos(2\vartheta t)}{\vartheta^{2}}\Bigg),
 \end{eqnarray}
  \begin{eqnarray} \label{Phy 16}
 C_{2}(n,t)=-\frac{1}{2}C_{1}(n,0)\frac{V_{1}^{2} \sin^{2}(\vartheta t)}{\vartheta^{2}},
\end{eqnarray}

 \begin{eqnarray} \label{Phy 17}
 C_{3}(n+1,t)&=&\frac{-i}{4}C_{1}(n,0)\Bigg(\sqrt{2}\sin(\sqrt{2} V_{1}t)+\frac{V_{1} \sin(2\vartheta t)}{\vartheta}\Bigg),
 \end{eqnarray}
  \begin{eqnarray} \label{Phy 18}
 C_{4}(n+1,t)&=&\frac{i}{4}C_{1}(n,0)\Bigg(\sqrt{2}\sin(\sqrt{2} V_{1}t)-\frac{V_{1} \sin(2\vartheta t)}{\vartheta}\Bigg),
  \end{eqnarray}
  \begin{eqnarray} \label{Phy 19}
C_{5}(n,t)&=&\frac{1}{4}C_{1}(n,0)\Bigg(-2\cos(\sqrt{2} V_{1} t)+\frac{V_{1}^{2}+2V_{2}^{2}+V_{1}^{2}\cos(2\vartheta t)}{\vartheta^{2}}\Bigg),
 \end{eqnarray}
   \begin{eqnarray} \label{Phy 20}
 C_{6}(n+2,t)=-C_{1}(n,0)\frac{V_{1} V_{2} \sin^{2}(\vartheta t)}{\vartheta^{2}},
  \end{eqnarray}
\end{subequations}
 where $\vartheta=\sqrt{V_{1}^{2}+V_{2}^{2}}$ and  $C_{1}(n,0)=\exp(-|\beta|^{2}/2)\beta^{n}/\sqrt{n!}$ determines the probability of the initial field state.

 \subsection{Two $\Xi$-type three-level atoms}
 In a similar manner, the state vector $|\psi_{2}(t)\rangle$ for two $\Xi$-type three-level atoms is given by:
       \begin{eqnarray}\label{Phy 16}
|\psi_{2}(t)\rangle_{\Xi}&=&\sum_{n=0}^{\infty} \Bigg(C_{1}(n,t)|1,1,n\rangle
+C_{2}(n+1,t) \big(|1,2,n+1\rangle+|2,1,n+1\rangle \big)\nonumber\\
&+&C_{3}(n+2,t)\big(|1,3,n+2\rangle+|3,1,n+2\rangle \big)\nonumber\\
&+&C_{4}(n+3,t)\big(|2,3,n+3\rangle+|3,2,n+3\rangle \big)\nonumber\\&+& C_{5}(n+2,t)|2,2,n+2\rangle+C_{6}(n+4,t)|3,3,n+4\rangle\Bigg).
 \end{eqnarray}
Following the same procedure as subsection \ref{Two $V$-type}, the values of the coefficients $C_{i}$ are obtained as below:
\begin{subequations}
 \begin{eqnarray} \label{Phy 17}
C_{1}(n,t)&=&\frac{C_{1}(n,0)}{x_{2} \eta} \Bigg((x_{2}-x_{4})\eta+(2V_{1}^{2} x_{2}-\beta_{2}^{2} x_{4} )\cos(\beta_{1}t)\nonumber\\
&-&(2V_{1}^{2}x_{2}-\beta_{1}^{2} x_{4})\cos(\beta_{2}t)\Bigg),
\end{eqnarray}
 \begin{eqnarray} \label{25}
C_{2}(n+1,t)&=&\frac{iC_{1}(n,0)}{2\beta_{1} \beta_{2}\eta V_{1}}
\Bigg((x_{4}-2\beta_{1}^{2} V_{1}^{2} )\beta_{2}\sin(\beta_{1}t)\nonumber\\&-&(x_{4}-2\beta_{2}^{2}V_{1}^{2})\beta_{1}
\sin(\beta_{2}t)\Bigg),
\end{eqnarray}
  \begin{eqnarray} \label{26}
C_{3}(n+2,t)&=&\frac{C_{1}(n,0)}{x_{2}\eta}
\Bigg(-x_{5}\eta-(\beta_{2}^{2}  x_{5}-V_{1} V_{2}  x_{2})\cos(\beta_{1}t)\nonumber\\&+&(\beta_{1}^{2} x_{5}-V_{1}V_{2} x_{2})\cos(\beta_{2}t)\Bigg),
\end{eqnarray}
\begin{equation} \label{27}
C_{4}(n+3,t)=\frac{-ix_{1}C_{1}(n,0)}{2V_{4}\eta}
\Bigg(\sin(\beta_{1}t)/\beta_{1}-\sin(\beta_{2}t)/\beta_{2}\Bigg),
\end{equation}
\begin{eqnarray} \label{28}
C_{5}(n+2,t)&=&\frac{2C_{1}(n,0)}{x_{2}\eta}
\Bigg(-x_{5}\eta-(\beta_{2}^{2}x_{5}-V_{1}V_{2}x_{2})\cos(\beta_{1}t)\nonumber\\&+&(\beta_{1}^{2}x_{5}-V_{1}V_{2}x_{2})\cos(\beta_{2}t)\Bigg),
\end{eqnarray}
\begin{equation} \label{29}
C_{6}(n+4,t)=\frac{x_{1}C_{1}(n,0)}{x_{2}\eta}
\Bigg(\eta-\beta_{1}^{2}\cos(\beta_{2}t)+\beta_{2}^{2}\cos(\beta_{1}t)\Bigg),
\end{equation}
\end{subequations}
with
\begin{eqnarray}\label{30}
x_{1}&=&6V_{1}V_{2}V_{3}V_{4},\hspace{.5cm}      x_{2}=6V_{1}^{2}V_{3}^{2}+4V_{1}^{2}V_{4}^{2}+6V_{2}^{2}V_{4}^{2},
\nonumber\\x_{3}&=&2(V_{1}^{2}+V_{4}^{2})+3(V_{2}^{2}+V_{3}^{2}),\hspace{0.5cm}
x_{4}=6V_{1}^{2}V_{3}^{2}+4V_{1}^{2}V_{4}^{2},
\nonumber\\x_{5}&=&2V_{1}V_{2}V_{4}^{2}, \hspace{1.cm} \eta=\sqrt{x_{3}^{2}-4x_{2}},
\nonumber\\ \beta_{1}&=&\sqrt{\frac{x_{3}+\eta}{2}}, \hspace{1.cm}\beta_{2}=\sqrt{\frac{x_{3}-\eta}{2}},\nonumber\\
V_{j}&=&V(n+j),\hspace{1.cm}j=1,2,3,4,\hspace{1.cm}V(n)=g\sqrt{n}.
\end{eqnarray}

\subsection{Two $\Lambda$-type three-level atoms}
Finally, the wave function of a system containing two $\Lambda$-type three-level atoms and a single-mode cavity field can be written as follows
   \begin{eqnarray} \label{Phy 31}
| \psi_{2}(t)\rangle_{\Lambda}&=&\sum_{n}^{\infty} \Bigg[C_{1}(n,t)|1,1,n \rangle +C_{2}(n+1,t)\big(|1,2,n+1 \rangle+|2,1,n+1 \rangle\big)\nonumber\\&+&C_{3}(n+1,t)\big(|1,3,n+1 \rangle+|3,1,n+1 \rangle\big)\nonumber\\&+&C_{4}(n+2,t)\big(|2,3,n+2 \rangle+|3,2,n+2 \rangle\big)\nonumber\\&+&C_{5}(n+2,t)|2,2,n+2 \rangle+C_{6}(n+2,t)|3,3,n+2\rangle)\Bigg],
  \end{eqnarray}
  where $C_{i}$ are the time-dependent probability amplitudes which must be obtained. Similarly, we arrive at
\begin{subequations}  \label{32}
\begin{eqnarray} \label{321}
C_{1}(n,t)&=&C_{1}(n,0)\frac{V_{2}^{2}+V_{1}^{2}\cos(2V_{3}t)}{V_{3}^{2}},
\end{eqnarray}
\begin{eqnarray} \label{322}
C_{2}(n+1,t)&=&C_{3}(n+1,t)=-iC_{1}(n,0)\frac{V_{1}\sin(2V_{3}t)}{2V_{3}},
\end{eqnarray}
\begin{eqnarray} \label{323}
C_{4}(n+2,t)&=&C_{5}(n+2,t)=C_{6}(n+2,t)\nonumber\\&=&-C_{1}(n,0)\frac{V_{1} V_{2} \sin^{2}(V_{3}t)}{V_{3}^{2}},
\end{eqnarray}
\end{subequations}
  where $V_{1}$ and $V_{2}$ have been previously defined in (\ref{30}).\\
   Here, it ought to be mentioned that, the relation
\begin{equation} \label{341}
 |\psi(t)\rangle_{V, \Xi, \Lambda}=\hat{D}(-\gamma)| \psi_{2}(t)\rangle_{V, \Xi, \Lambda}, \hspace{2cm} \gamma=\frac{\lambda}{g}
 \end{equation}
allows us to write the solution of DJCM (Eq. (\ref{Phy 1})) for two $V$-, $\Xi$- and $\Lambda$-type three-level atoms, explicitly. For example for two $\Lambda$-type atoms the state vector is in the following form:
  \begin{eqnarray} \label{Phy 311}
| \psi(t)\rangle_{\Lambda}&=&\sum_{n}^{\infty} \Bigg[C_{1}(n,t)|1,1\rangle |-\gamma;n \rangle \nonumber\\&+&C_{2}(n+1,t)\Big(|1,2\rangle|-\gamma;n+1 \rangle+|2,1\rangle|-\gamma;n+1 \rangle \Big)\nonumber\\&+&C_{3}(n+1,t)\Big(|1,3\rangle|-\gamma;n+1 \rangle+|3,1\rangle|-\gamma;n+1 \rangle\Big)\nonumber\\&+&C_{4}(n+2,t) \Big(|2,3\rangle|-\gamma;n+2 \rangle+|3,2\rangle|-\gamma;n+2 \rangle\Big)\nonumber\\&+&C_{5}(n+2,t)|2,2\rangle|-\gamma;n+2 \rangle\nonumber\\&+&C_{6}(n+2,t)|3,3\rangle|-\gamma;n+2 \rangle)\Bigg],
  \end{eqnarray}
  where $|-\gamma;n+j \rangle$, $j=0,1,2$, are the displaced number states.
  Similar expressions can be simply obtained for other two types of atoms.
  Therefore, our proposed model can also be considered as a novel scheme for the generation of displaced number states \cite{Oliveira1990}. By considering similar approach for two other cases ($V$- and $\Xi$-type three-level atoms), the displaced number states can be generated, too.
  Anyway, we are now ready to study the nonclassical properties of three different types of the atom-field system by emphasizing on their entanglement properties. For this purpose we will pay attention to entanglement, photon statistics and quadrature field squeezing.
   \section{Entanglement} \label{Entanglement}
Entanglement is the noticeable feature of quantum states which demonstrates correlations that cannot be  classically accounted. The first investigations of entanglement date back to 1935, focusing on surprising consequences of the quantum description of nature \cite{Einstein1935,Schrodinger1935}. Entangled qubits are an urgent resource in many quantum information applications such as quantum computation and quantum communication \cite{Bennett.DiVincenzo2000}, quantum metrology \cite{Giovannetti2004}, quantum cryptography \cite{Jennewein.etal2000}, quantum teleportation \cite{Ganguly.etal2011} and other applications in quantum technology \cite{Guehne2009,Horodecki2009}. Recently, much attention has been paid to the generation of quantum entangled states. A well-known source for the generation of such states is the atom-field interaction process, using different models of interaction. Accordingly, the most interesting aspects of the JCM and its generalizations, which has received much attention, is the possible existing of entanglement between different subsystems \cite{Akhtarshenas2007,Sainz2007,Paula2014}.
So, we now pay attention to the evaluation of the entanglement dynamics of the obtained states.
To achieve this goal, several suitable measures of DEM  such as von Neumann entropy \cite{Vedral1997}, entanglement of formation \cite{Wootters1998}, concurrence \cite{Wootters1998,Wootters2001} and negativity \cite{Vidal2002} have been proposed. In this section we apply von Neumann entropy and negativity to investigate entanglement dynamics of atom-field and atom-atom, respectively.
It ought to be mentioned that, while the von Neumann entropy is a good measure for the atom-field entanglement, this measure is not appropriate for the calculation of the DEM between the two atoms (this is due to the fact that in this case, the system (the two atoms) is a mixed state). Also, the evaluation of DEM between ``two atoms'' and ``field'' by negativity is not an easy task (its complexity arises from the fact that to achieve this purpose we are left with the infinite dimensional Hilbert space related to the fields).
 \subsection{von Neumann entropy} \label{von Neumann entropy}
We have assumed that two three-level atoms and the coherent field are initially in a pure state. So, the considered  systems can be regarded as the bipartite systems, consisting of two three-level atoms as the first subsystem  and the radiation field as the second subsystem. For such systems, the von Neumann entropy is a suitable measure to obtain the DEM between subsystems \cite{Phoenix1988}.
Quantum mechanically, the von Neumann entropy for a quantum system with the density operator $\rho$  is defined as $S=-\mathrm{Tr}(\rho \ln \rho)$.  If $\rho$ describes a pure state, then $S=0$, and if it represents a mixed state then $S\neq0$.
 For the considered atom-field systems, the entropy of the field or equivalently the atoms is a good measure to realize the amount of entanglement; higher (lower) entropy means the greater (smaller) DEM. Before obtaining the reduced entropy of the field and the atom, it is worth to pay attention to the important theorem of Araki and Leib \cite{Araki.Lieb1970}. According to this theorem, for any bipartite quantum system, the system and subsystem entropies at any time $t$ are bounded by the triangle inequality $|S_{A}(t)-S_{F}(t)|\leq S\leq|S_{A}(t)+S_{F}(t)|$,
where $S_{A}$ and $S_{F}$ represent the entropies of the atom and field, respectively, and the total entropy of the atom-field system is denoted by $S$. One immediate consequence of this inequality is that, if at the initial time the whole system is prepared in a pure state, the total entropy of the system is zero and remains constant when the involved subsystems are isolated from their environment. This implies that, if the system is initially in the pure state, $ S_{AF}=0 $, at any time $ t > 0$ the field and atomic entropies are equal \cite{Barnett.Phoenix1991}. Therefore, instead of the calculation of the field entropy, we can evaluate the entropy of the atoms. According to the von Neumann entropy, the entropies of the atom and field, when treated as a separate system, are defined through the corresponding reduced density operators as
\begin{equation} \label{Von 2}
 S_{A(F)}(t) = - \mathrm{Tr}_{A(F)}(\hat{\rho}_{A(F)}(t)\ln\hat{\rho}_{A(F)}(t)),
\end{equation}
with $\hat{\rho}_{A(F)}(t)=\mathrm{Tr}_{F(A)}(|\psi(t)\rangle \langle\psi(t)|)$, as the reduced density operator of the atoms (field). Now, we turn our attention to discuss the DEM between two atoms and quantized field for three different systems through the von Neumann entropy. We assume that the considering systems start from a pure state, so the entropy of the field/atom may be expressed by the following relation
\begin{equation} \label{25}
\mathrm{DEM}(t) = S_{F}(t) = S_{A}(t)=- \sum_{i=1}^{9}\xi_{i} \ln\xi_{i},
\end{equation}
where $\xi_{i}$ denote the eigenvalues of the reduced density matrix of the atoms, which can be obtained numerically for $V$- and $\Xi$-type, but analytically for $\Lambda$-type configuration. Indeed, the density matrix of the two-atom system in the $\Lambda$-type configuration has rank no larger than three with the associated eigenvalues evaluated, analytically, by the Cardano's method as \cite{Childs2009}
\begin{eqnarray}
\xi_{j} &=& -\frac{1}{3} \varrho_{1}+\frac{2}{3}\sqrt{\varrho_{1}^{2} - 3 \varrho_{2}} \cos \left(\varpi+\frac{2}{3}(j-1)\pi \right),\;\;\;\;\;\;\;  j =1,2,3, \nonumber \\
\xi_{j} &=& 0,\;\;\;\;\;\;\;  j = 4,5,6,7,8,9,
 \end{eqnarray}
with
\begin{equation} \label{27}
\varpi=\frac{1}{3}\cos^{-1}\left[ \frac{9 \varrho_{1}\varrho_{2}-2 \varrho_{1}^{3} - 27 \varrho_{3}}{2( \varrho_{1}^{2} - 3\varrho_{2})^{3/2}}\right],
\end{equation}
\begin{subequations}
and
\begin{eqnarray}\label{271}
\varrho_{1} &=& -\rho_{11}-4(\rho_{22}+\rho_{44}), \label{28-1}
\end{eqnarray}
\begin{eqnarray}\label{272}
\varrho_{2} &=& -4(\rho_{12}\rho_{21}+\rho_{14}\rho_{41}+4\rho_{24}\rho_{42})+4\rho_{11}(\rho_{22}+\rho_{44})+16\rho_{22}\rho_{44},
\end{eqnarray}
\begin{eqnarray}\label{273}
\varrho_{3} &=& 16\rho_{14}(\rho_{22}\rho_{41}-\rho_{21}\rho_{42})+16\rho_{12}(\rho_{21}\rho_{44}-\rho_{24}\rho_{41}) \nonumber\\&+& 16\rho_{11}(\rho_{24}\rho_{42}
-\rho_{22}\rho_{44}).
\end{eqnarray}
\end{subequations}
Moreover, the matrix elements of the atomic density operator are as follows:
\begin{eqnarray}\label{281}
\rho_{ij}(t)&=&\sum_{n=0}^{+\infty} C_{i}(n,t)C_{j}^{\ast}(n,t), \hspace{.5cm}
i,j = 1,2, \cdots 6.
\end{eqnarray}
where $C_{i}(n,t)$ have been derived in \ref{32}.
 Figure 2 shows the evolution of the field entropy against the scaled time $gt$ for initial mean photon number fixed at $|\alpha|^2=25$ and two atoms prepared initially in the higher exited state. Frames 1, 2, and 3 respectively concern with the two $V$-, $\Xi$- and $\Lambda$-types three-level atoms. Also, panels (a) in this figure are plotted in the absence of classical field ($\gamma=0$) and panels (b) and (c) show the effect of the driving external field on the behaviour of von Neumann entropy ($\gamma=2$ and $\gamma=6$). It can be obviously seen from this figure that for three different configuration of three-level atoms, the field entropy gets the maximum value of entanglement after the onset of the interaction. In addition, it can be observed from frame 1 that, entering the classical field together with increasing the related parameter may lead to the shift of the maxima amounts of the DEM when the time proceeds. Comparing the panels 1 and 2 indicates that the temporal behaviour of  two $\Xi$-type three-level atoms is qualitatively the same as two $V$-type ones. For the case that we deal with $\Lambda$-type three-level atoms, by considering the effect of external classical field, it seems that the temporal behaviour of the DEM behaves oscillatory specially in the case $\gamma = 6$. Finally, by focusing on the effect
of external classical field on the behaviour of the von Neumann entropy, it is found that the existence of the classical field may increase the maximum values of DEM between two atoms and field in three considered systems.
\subsection{Negativity} \label{Negativity}
In the present subsection, we apply the negativity measure for the investigation of the DEM between two atoms. Among all entanglement measures, negativity surely is the best known and most popular instrumentation to specify bipartite quantum correlations \cite{Eltschka2013}. It is easily evaluated for arbitrary states of a composite system and so can be applied to study entanglement in many different situations. Historically, this quantity, which is based on the Peres-Horodecki criterion for the separability of a state \cite{Peres1996,Horodecki1996}, was first used by Zyczkowski et al \cite{ifmmodedotZelse.Zfiyczkowski1998} and subsequently introduced by Vidal and Werner as a new entanglement measure \cite{Vidal2002}. The positive partial transpose (PPT) is a necessary and sufficient condition for separability of $2\times2$ and $2\times3$-dimensional mixed states \cite{Peres1996,Horodecki1996}, arbitrary $d \times d'$ pure state, all Gaussian states of $1\times n$  mode continuous variable systems \cite{Simon2000,Werner.Wolf2001} and also any state $\rho$ which is supported on $d  \times d'$ systems (d $\leq d'$) and with rank $r(\rho)\leq d'$ \cite{Horodecki2000}, but it is only sufficient, in general, for other systems. The negativity for a bipartite quantum system with $d \times d'$ ($d \leq d'$)-dimensional Hilbert space $ \mathcal{H}_{A} \otimes \mathcal{H}_{B}$ described by the density matrix $\hat{\rho}$, is given by
$(d\leq d')$
  \begin{equation} \label{N2}
\mathcal{N}(\rho)=\frac{|| \hat{\rho}^{T_{B}}||_{1} - 1}{d - 1},
\end{equation}
where $\hat{\rho}^{T_{B}}$  is the partial transpose of the state $\hat{\rho}$ with respect to subsystem $B$ and $||\hat{M}||_{1}=\mathrm{Tr}\sqrt{\hat{M}^{\dag}\hat{M}}$ is the trace class norm of the operator $\hat{M}$, which reduces to the sum of the absolute value of the eigenvalues of $\hat{M}$, when $\hat{M}$ is Hermitian.
The matrix $\hat{\rho}$ is a positive operator with trace one, i.e. $\mathrm{Tr}(\hat{\rho})=1$. Also, for the partial transpose of this matrix we have $\mathrm{Tr}(\hat{\rho}^{T_{B}})=1$, too. Since the partial transpose of density operator might have the negative eigenvalues, the trace norm of $\hat{\rho}^{T_{B}}$ can be written in the following form
\begin{equation}\label{332}
||\hat{\rho}^{T_{B}}||_{1}=\sum_{i}|\mu_{i}|=\sum_{i}\mu_{i}-2\sum_{i}\mu^{neg}_{i}=1-2\sum_{i}\mu^{neg}_{i},
\end{equation}
where $\mu_{i}$ and $\mu^{neg}_{i}$ correspond to the positive and negative eigenvalues of $\hat{\rho}^{T_{B}}$, respectively. Here, we are going to evaluate the DEM between two three-level atoms ($(3\times 3)$-dimensional Hilbert space) in the considered systems.
\\In Fig. 3, we have plotted the time evolution of the negativity as a function of the dimensionless time $gt$ for the same parameters as in Fig. 2. One can see from this figure that the temporal behavior of the negativity in various conditions represents irregular oscillations between minima and maxima values. Frame 1 of this figure shows the DEM between two $V$-type three-level atoms with and without the external classical field. According to these figures, it may be seen that by increasing the amplitude of the external classical field, the maxima values of negativity are revealed with time delay. For two $\Xi$-type three-level atoms, the outstanding effect of the classical field is decrement the sustainment time of the maxima values of the negativity. The same behavior can be seen in frame 3 for two $\Lambda$-type three-level atoms. Finally, by considering the presented results depicted in Fig. 3, it is found that the DEM between $\Xi$-type three-level atoms in the presence or absence of the classical field is greater than two other configurations. Also, for different types of three-level atoms the negativity can be managed by the driving external classical field. On the other hand, the addition of the classical field causes a slight increase in the maximum values of the DEM between two atoms.
  \section{Photon statistics: the Mandel parameter}\label{Photon statistics: the Mandel parameter}
Sub-Poissonian statistics is a striking feature of  nonclassical states. To investigate the statistical properties of any system the Mandel parameter is a suitable measure \cite{Mandel1979}. This parameter has been defined as the following form
 \begin{equation}\label{35}
Q(t)=\frac{\langle n^{2}\rangle-\langle n\rangle^2}{\langle n\rangle}-1.
\end{equation}
This quantity is positive, zero and negative when the statistics is super-Poissonian, Poissonian and sub-Poissonian, respectively. The sub-Poissonian statistics is a sufficient but not necessary condition for nonclassicality of the field \cite{Agarwal1992a}.
For our considered systems we have:
\begin{eqnarray}\label{36}
\langle n\rangle &=&\langle\psi(t)| \hat{a}^{\dag} \hat{a}|\psi(t)\rangle=
\langle\psi_{2}(t)| \hat{D}(\gamma)\hat{a}^{\dag} \hat{a}\hat{D}^{\dag}(\gamma)|\psi_{2}(t)\rangle\nonumber\\&=&\langle\psi_{2}(t)| (\hat{a}^{\dag}-\gamma) (\hat{a}-\gamma)|\psi_{2}(t)\rangle=\gamma^{2}+\langle\psi_{2}(t)|\hat{a}^{\dag}\hat{a}|\psi_{2}(t)\rangle
\nonumber\\&-&\gamma(\langle\psi_{2}(t)|\hat{a}|\psi_{2}(t)\rangle+\langle\psi_{2}(t)|\hat{a}^{\dag}|\psi_{2}(t)\rangle),
\end{eqnarray}
\begin{eqnarray}\label{37}
\langle n^{2}\rangle &=&\langle\psi(t)|( \hat{a}^{\dag} \hat{a})^{2}|\psi(t)\rangle=\langle n\rangle +
\langle\psi_{2}(t)| (\hat{a}^{\dag}-\gamma)^{2} (\hat{a}-\gamma)^{2}|\psi_{2}(t)\rangle.
\end{eqnarray}
To visualize the effect of the external classical field on the photon statistics for the different three configurations, we have plotted the Mandel parameter against the scaled time $gt$ for different values of the coupling parameter ratio $\gamma$ in Fig. 4. In the absence of classical field, the oscillations of the Mandel parameter in three configurations show collapse and revival phenomena as would be expected. Also, in these cases the Mandel parameter varies between positive and negative values, which means that the photons display super- or sub-Poissonian statistics for different intervals of times, alternatively. Due to the presence of external field,
we observe that the Mandel parameter possesses a periodic behavior in the positive region at the most of times. This means that in these times, the entire atom-field state of the considered systems has a super-Poissonian statistics. Altogether, in the presence of classical field and only in the beginnings time of the interaction the field has sub-Poissonian statistics. Increasing the value of the classical field coupling parameter leads to the decrease in the time average of the sub-Poissonian statistics. Finally, it is to be noted that the negative value of Mandel parameter in $\Xi$-type is greater than its counterparts in both $V$- and $\Lambda$-types.
  \section{Normal squeezing of the field}\label{Squeezing}
In the present section we focus on the squeezing phenomenon. This parameter is described by decreasing the quantum fluctuations in one of the field quadratures with the price of an increase in the corresponding conjugate quadrature, such that the Heisenberg uncertainty principle is not violated. The squeezed light has attractive applications in optical communication networks and gravitational wave detection \cite{Bachor2004,Kimble1987}. The squeezing parameter has been defined in various ways such as normal squeezing, amplitude-squared squeezing \cite{Hillery1987}, higher-order squeezing \cite{Collett1984} and principal squeezing \cite{Perina1991}. However, in the present section we pay attention to the normal squeezing. \\
To investigate the normal squeezing of the field, we introduce two quadrature field operators $\hat{x}=(\hat{a}+\hat{a}^{\dagger})/2$ and $\hat{y}=(\hat{a}-\hat{a}^{\dagger})/2i$. The system then would possess squeezing if one of the quadratures convinces the inequality $\langle (\Delta\hat{x})^{2} \rangle<0.25$ or $\langle (\Delta\hat{y})^{2} \rangle<0.25$ where $\langle (\Delta\hat{x_{i}})^{2} \rangle=\langle \hat{x_{i}}^{2} \rangle-\langle
\hat{x_{i}}  \rangle^{2}$, $x_{i}=x$ and $y$. Equivalently, if we define $S_{x}=4\langle (\Delta\hat{x})^{2} \rangle-1$ and $S_{y}=4\langle (\Delta\hat{y})^{2} \rangle-1$, squeezing occurs in $\hat{x}$ ($\hat{y}$) component if $-1<S_{x}<0$ ($-1<S_{y}<0$). These parameters can be rewritten as follows:
\begin{equation}\label{25}
S_{x} =2 \langle \hat{a}^\dagger \hat{a} \rangle + 2 \Re \langle \hat{a}^{2} \rangle  -4 (\Re \langle \hat{a} \rangle )^{2},
\end{equation}
\begin{equation}\label{26}
S_{y} = 2 \langle \hat{a}^\dagger \hat{a}\rangle -  2 \Re \langle \hat{a}^{2} \rangle  -4 (\Im \langle \hat{a} \rangle)^{2},
\end{equation}
where $\langle \hat{a}^\dagger \hat{a} \rangle$ has been previously defined in (\ref{36}) and $\langle \hat{a}^{m} \rangle$ can be obtained in the following form
\begin{eqnarray}\label{371}
\langle  \hat{a}^{m}\rangle &=&\langle\psi(t)|\hat{a}^{m}|\psi(t)\rangle=\langle\psi_{2}(t)| (\hat{a}-\gamma)^m|\psi_{2}(t)\rangle.
\end{eqnarray}
Our results presented in Fig. 5 indicate the time evolution of the quadrature squeezing parameters $S_{x}$ and $S_{y}$ against the scaled time $gt$ for different types of three-level atoms as well as various values of the coupling parameter ratio. We can see from this figure that for all cases, the squeezing exists in the $x$ quadrature and no squeezing is occurred in the $y$ quadrature. Meanwhile, squeezing arises in $x$ quadrature only at the beginning of the occurrence of the atom-field interaction. Furthermore, a comparison between frame (a) and frames (b,c) of different configurations of three-level atoms shows that the amounts of this nonclassical effect in $x$ diminishes by increasing $\gamma$. Also, our conclusions represent that, the squeezing in $x$ component in the $\Xi$-configuration is stronger than for the other two configurations.
  \section{Conclusion}\label{Conclusion}
Due to the importance of the three-level atoms as well as the presence of the driving classical field in the atom-field interactions, in this paper, we have outlined two identical three-level atoms (in $V$, $\Xi$ and $\Lambda$ configurations) interacting with a quantized single-mode field assisted by an external classical field. It is shown that the introduced system can be transformed to the usual form of the generalized JCM by using two appropriate unitary transformations. Therefore, we could solve the dynamical problem and find the explicit form of the entangled state vector of the three different considered atom-field systems analytically, by considering the atoms initially in the higher exited state and the quantized field in the coherent state. Next, at first, the quantum entanglement between the subsystems of the generated states are computationally evaluated by using the approach of von Neumann entropy (to study the DEM between two atoms and quantized field) and negativity (to investigate the DEM between two atoms). Then, the quantum statistics and quadrature squeezing of the obtained states have been numerically investigated. In each case, we studied the effect of the external classical field on the mentioned physical quantities. The main results of the paper are listed in what follows.
\begin{enumerate}
   \item
Generally, entering the classical field on the interaction together with increasing the related parameter may lead to the shift of the maxima amounts of the field entropy for two $V$- and $\Xi$-type three-level. Increasing the value of $\gamma$ leads to the oscillatory behaviour of the von Neumann entropy for $\Lambda$-type configuration.
  \item
Maximum values of the DEM between two atoms for $\Xi$-configuration are larger than those for other two configurations. Also, the DEM between two atoms depend on the driving classical field.
 \item
Since the rank of the reduced density operator of the atomic system containing two $\Lambda$-type three-level atoms is three, it is deduced that the PPT is necessary and sufficient condition for separability and in this case, the negativity fully captures the entanglement of this system. It turns out that under conditions considered in this work, the  atom-atom entanglement generated in the $\Lambda$-type configuration is distillable.
\item
The increase in the external classical field is associated with a slight increment in the maximum values of the DEM between different subsystems.
  \item
The complete (partial) collapse and revival, as purely quantum mechanical features, are observed in Mandel parameter in the absence of the classical field. Also, in the presence of classical field the quantized cavity mode is super-Poissonian after certain interaction time.
  \item
The numerical results indicate that, no squeezing is seen in $y$ component and squeezing occurs in $x$ quadrature in very short time intervals in the beginning of the interaction. Also, it is obviously seen that the profundity of squeezing in these regions is decreased by the increment of the amplitude of the classical field.
  \item
  It is illustrated that the amount of considered entanglement criteria as well as quantum statistics and squeezing can be tuned by applying the external classical field appropriately via parameter $\gamma$. This, however, is more clear in quantum statistics and squeezing as compared with entanglement criteria.
  \item
 As a marginal result of the paper, we would like to state that, in the appropriate conditions our proposal can be used for the generation of displacement number states.
 \end{enumerate}
 At the end of this paper, we mention that this study can be accomplished by considering spontaneous emissions and other decoherence processes. We hope to report this work in the near future elsewhere.
\section*{References}
\providecommand{\newblock}{}

\newpage

\end{document}